\documentclass[]{aa}
\usepackage{natbib}
\usepackage{graphicx}
\usepackage{txfonts}
\usepackage{color}
\def\fH2{\mbox{f$_\HH$}}

\def\EBV{\mbox{E$_{\rm B-V}$}}

\def\AV{\mbox{A$_{\rm V}$}}

\def\nH2{\mbox{${\rm n}_\HH}$}
\def\Rgal{\mbox{R$_{\rm gal}$}}
\def\pccc{~{\rm cm}^{-3}} 
 
\def\pcc{~{\rm cm}^{-2}}

\def\Tsub#1 {\mbox{${\rm T}_{\rm #1}$}}
\def\TK  {\Tsub K }

\def\Texc {\Tsub ex }

\def\arcsec{\mbox{$^{\prime\prime}$}} \def\arcmin{\mbox{$^{\prime}$}}

\def\degr{$^{\rm o}$}
\def\p{\mbox{$^+$}}

\def\h13cop{\mbox{{H$^{13}$CO\p}}}

\def\C3H{\mbox{C$_3$H}}
\def\c3h2{\mbox{C$_3$H$_2$}}
\def\cc3h2{\mbox{{\it c}-C$_3$H$_2$}}
 
 \def\R0{R$_0$}
\def\G0{\mbox{G$_0$}}

\def\ddeg{{}^\circ\kern-.1em}

\def \kms{\mbox{km\,s$^{-1}$}}

\def\E#1 {$10^{#1}$}
\def\E#1 {E{#1}}
\def\P#1,{$\nH2\TK~=~#1\times~10^4\pccc$~K}
\def\ec#1,#2,#3,{#1\,(#2)\E{#3}}

\def\H3{\mbox{H$_3$}}

\def\RH2{\mbox{R$_{\rm G}$}}
\def\g13{\mbox{g$_{13}$}} 

\def\cc3h{\mbox{{\it c}-\C3H}}
\def\lc3h{\mbox{{\it l}-\C3H}}

\newcommand{\emm}[1]{\ensuremath{#1}}   
\newcommand{\emr}[1]{\emm{\mathrm{#1}}} 

\newcommand{\hcop}{\emr{HCO^+}} 
 
\newcommand{\HH}{\emr{H_2}}

\newcommand{\W}[1]{\emm{{\rm W}_\emr{#1}}}
\newcommand{\WCO}{\W{CO}}

\newcommand{\Msun}{\emm{M_\odot}}

\title{ALMA hints at the existence of an unseen  reservoir of diffuse molecular 
 gas in the Galactic bulge}

\author{M. Gerin\inst{1} and H. Liszt\inst{2}}
\institute{ 
LERMA, Observatoire de Paris,  PSL Research University, CNRS,
Sorbonne Universit\'es, UPMC Univ. Paris 06, Ecole Normale
  Sup\'erieure, F-75005 Paris, France
\email{maryvonne.gerin@ens.fr}
\and       National Radio Astronomy Observatory,
           520 Edgemont Road,
           Charlottesville, VA,
           USA 22903
\email{hliszt@nrao.edu}
}
\begin{document}

\date{received \today}%
\offprints{H. S. Liszt}%
\mail{hliszt@nrao.edu}%
%
\abstract
{}
{We aim to understand  the unexpected presence of mm-wave molecular absorption at 
-200 \kms $< {\rm v} < -140$ \kms\ in a direction that is well away from regions
 of the Galactic bulge where CO emission at such velocities is prominent.}
{We compared 89 GHz Cycle 2 ALMA absorption spectra 
 of \hcop, HCN, and HNC toward the extragalactic continuum source 
  B1741-312 at l=-2.14\degr, b=-1.00\degr\ with existing CO, H I, and dust 
 emission and absorption measurements.   We placed the atomic and molecular gas 
 in the bulge and disk using circular and non-circular galactic kinematics, 
 deriving N(H I) from a combination of 21cm emission and absorption and we derive
 N(\HH) from scaling of the \hcop\ absorption.  We then inverted the variation of 
near-IR reddening 
 E(J-K) with distance modulus and scale E(J-K) to a total gas column density N(H)
 that may be compared to N(H I) and N(\HH). }
 {At galactocentric radii \Rgal\ $>$ 1.5 kpc, conventional measures such 
 as the standard CO-\HH\ conversion factor and locally observed N(\hcop)/N(\HH) 
 ratio separately imply that H I and \HH\ contribute about equally to N(H), and 
 the gas-derived N(H) values are in broad  agreement with those derived from E(J-K).  
 Within the Galactic bulge at \Rgal $<$ 1.5 kpc, H I 
 contributes less than 10\% of the material inferred from E(J-K), so
that the  molecular absorption detected here is needed
 to understand the extinction.}
 {}

\keywords{ interstellar medium -- abundances; B1741-312 }

\authorrunning{Gerin, Liszt} \titlerunning{Diffuse molecular gas in the Galactic bulge}

\maketitle{}

%

\section{Introduction}

The distribution of molecular gas in the Galactic nucleus has been heavily studied 
 in the so-called central molecular zone (CMZ) \citep{MorSer96} that
extends 
$\pm$ 1.2\degr\ in Galactic longitude or $\pm$ 180 pc projected distance 
at the IAU standard distance R$_0$ = 8.5 kpc.   The molecular component 
outside the CMZ, but within the bulge, is far less well understood owing to 
limited latitude coverage and sparse sampling of the CO survey 
\citep{BitAlv+97} and to uncertainty in the conversion from CO brightness to 
\HH\ column density \citep{BolWol+13}.

In the Galactic bulge, the atomic gas has been fully if somewhat coarsely mapped 
\citep{BurLis83,HI4PI}, and the mass of H I, $3\times 10^7$ \Msun\, is 
comparable to that 
of the molecular gas within the much more compact CMZ.  The atomic gas shows
a broad range of phenomena associated with large-scale gas flow in the Galactic 
bar outside the CMZ \citep{Fux99} but only a small portion 
of that is seen in CO and other molecules \citep{RodCom+06}.
Molecular emission is present in a few very broad-lined features associated
with the shredding and intake of molecular gas at the standing dust lane
shock in the Galactic bar, and CO is detected in the terminal-velocity feature 
associated with inward gas flow along the dust lane \citep{Lis06Shred,Lis08Multi}.
Even CO emission is largely absent from most of the velocity range observed in 
H I, however, presumably because of limited sensitivity to weak, broad emission.  
This led to the suggestion \citep{LisBur78} that a large reservoir of molecular 
gas had yet to be observed in the bulge.

Some of this gas may have been serendipitously observed in absorption in 
an ALMA experiment studying the chemistry of CF\p\ and HOC\p\ in local 
diffuse molecular gas, and this paper explains why that might be the case.  
The structure of this work is as follows.  In Sect. 2 we describe the new and
existing observational material that is discussed.  In Sect. 3 we present
our observational results, from which we derive column densities of diffuse
atomic and molecular gas in the Galactic disk and bulge, and compare these
with gas column densities implied by the variation of near-IR dust reddening 
E(J-K) with distance modulus.  Section 4 is a summary and discussion.

\section{Observations and data reduction}

\begin{figure}
\includegraphics[height=9.6cm]{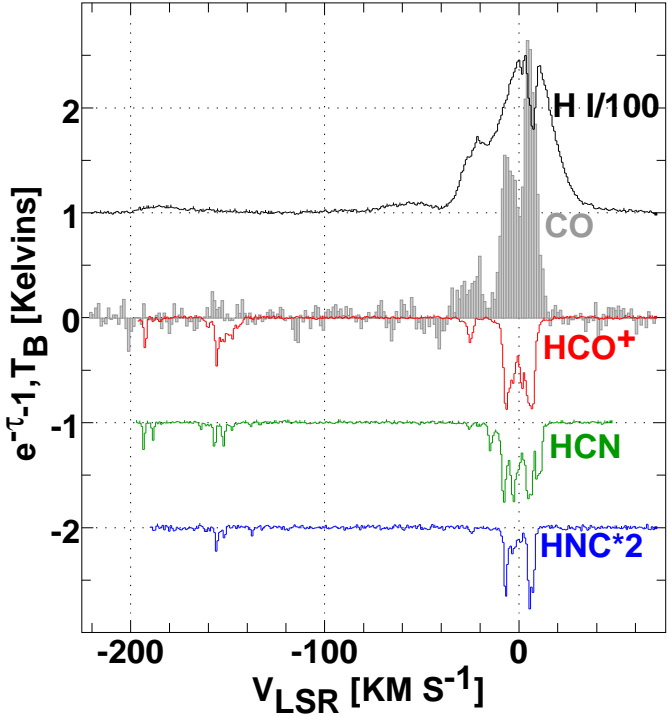}
  \caption[]{SGPS H I, Southern Mini CO, and ALMA \hcop, HCN, and HNC
   profiles. The strongest CO emission feature is the counterpart of
  a strong and very widespread H I self-absorption feature.}
\end{figure}

\begin{figure}
\includegraphics[height=9.6cm]{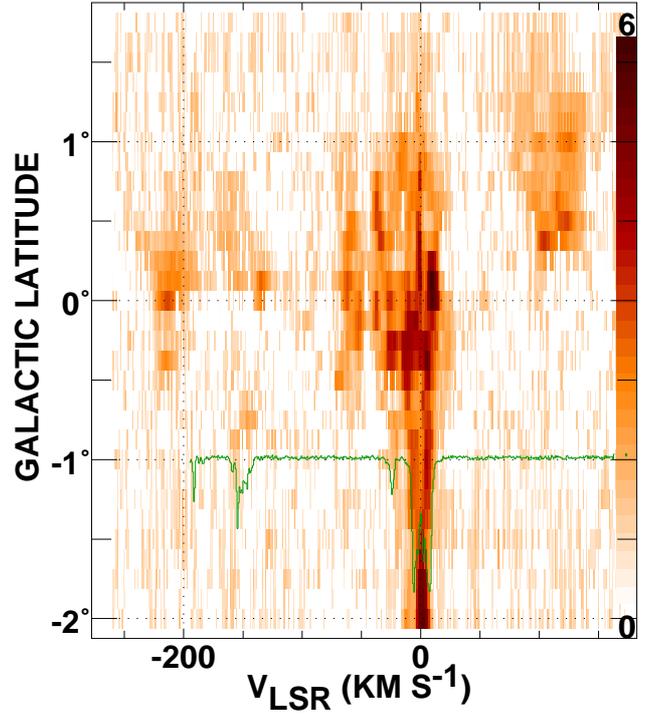}
  \caption[]{Latitude-velocity diagram of CO emission at $l =-2.125$\degr\ 
 with the ALMA \hcop\ absorption profile superposed such that its 0-level 
 is at the latitude of B1741-312.}
\end{figure}

\subsection{ALMA absorption measurements}

We observed \hcop, HCN, and HNC in absorption near 89 GHz in ALMA Cycle 2 project 
2013.1.01194.S, whose scientific goal was a comparative study of the chemistries 
of CF\p\ and HOC\p\ with respect to HF and \HH O in diffuse molecular gas.  The blazar 
background source B1741-312 at $l=-2.1366$\degr,  $b=-0.9968$\degr\ was observed 
in nine spectral line windows, three of which covered the J=1-0 transitions of 
HCN, \hcop, and HNC at 88.631 GHz, 89.188 GHz, and 90.663 GHz, respectively, as
discussed here.  The appearance of the HCN absorption is complicated by the inherent 
hyperfine splitting into three components with local thermal
equilibrium (LTE) optical depths in the ratio 
1:5:3 in velocity order.  Observations of HOC\p, CF\p, C$_3$H\p,\ and c-C$_3$H in 
the nearby gas will be discussed 
in a forthcoming paper.
  
Although the intended subject of study toward B1741-312 was relatively local gas 
near 0-velocity, the spectral setup was held in common with those used for Sgr A 
and Sgr B in the same science goal, therefore
 extending almost to -200 \kms:  the lowest-frequency transition, that of HCN, 
extends the farthest to negative velocity, detecting (with \hcop) a feature at -190 
 \kms\ that  was narrowly missed in the highest frequency spectral window, 
that for HNC.  
The channel spacing of the ALMA data presented here is 244 kHz, or 
approximately 0.8 \kms, and the resolution is twice as high.  

The continuum flux was 0.59 Jy and the source was observed for 720s, leading 
to channel-channel rms optical depth fluctuations of 0.0062 at zero optical 
depth.  The spectra were reduced using the standard ALMA pipeline in 
January and February 2016 after the data had been delivered.  Spectra were 
extracted from the final data cubes at the pixel of peak continuum flux.

 The velocities  of these and all other data discussed here are measured 
with respect to the kinematic definition of the Local Standard of Rest (LSR).

\subsection{Other data: H I and CO emission}

We present CO J=1-0 emission profiles from the Southern Mini-Telescope CO survey 
of \cite{BitAlv+97} at 8\arcmin\ spatial resolution on a 1/8\degr\ grid and 
H I emission from the Southern Galactic Plane Survey (SGPS) survey data of 
\cite{McCDic+05} at 145\arcsec\ spatial resolution, which we used in conjunction 
with the H I absorption line profile of \cite{DicKul+83} to derive an 
optical-depth corrected N(H I) as

$$N(H~I) = 1.823\times 10^{18}\pcc \int \frac{\tau(v)T_B(v)}{1-\exp{(-\tau(v))}} dv,
\eqno(1) $$

where $\tau$ and T$_B$ are the optical depth and brightness temperature
and the units of velocity are \kms.

\subsection{Extinction}

We also use data at 6\arcmin\ resolution from the 3D IR extinction map of the bulge 
of \cite{SchChe+14}, extracting the variation of E(J-K) along the line of sight at 
(l,b) = (-2.1\degr,-1.0\degr) and using the relationships A$_{\rm K}$/\AV\ = 0.11,
A$_{\rm K}$/E(J-K) = 0.689, or \AV\ = 6.26 E(J-K) and 
 N(H) $= 1.171\times 10^{22}\pcc$ E(J-K), where N(H) is the total column density
of H-nuclei in atomic and molecular form,  N(H) = N(H I)+2N(\HH).
 
\section{Observational results}

Atomic and molecular absorption and emission spectra toward and around B1741-312 are 
shown in Fig. 1.  Strong molecular absorption lines are seen from local gas around 
0-velocity, and the molecular feature at v = 7 \kms\ corresponds to a strong H I 
self-absorption feature originally discovered by \cite{Hee55} that provided the
first evidence of inhomogeneity in the temperature structure of the interstellar
H I gas.  
The high column density of local gas seen around 
0-velocity dominates the total absorption and emission in this direction, but
molecular absorption also appears around -25, -150, and -190 \kms.  The spectrum of
 HNC, lying at the highest rest frequency, did not quite extend out to the narrow
absorption near -190 \kms.

Molecular absorption at -190 \kms\ from the Galactic bulge is narrow in \hcop\ and 
HCN, as perhaps arising from a single isolated cloud.  The absorption around -150 \kms\ is 
wider and more complex, with two kinematic sub-components separated by 6 \kms\ in HNC 
 and 
more continuous absorption over 20 \kms\ in \hcop.  The absorption depth 
ratios of the three species and the somewhat greater ubiquity of \hcop\ in the high
negative-velocity bulge features are typical of molecular absorption from diffuse 
molecular gas in the interstellar medium (ISM) near the Sun \citep{LisLuc01}, or the 
Galactic disk 
\citep{GodFal+10}.

H I emission is apparent around -190 \kms\ toward B1741-312 in Fig. 1, but
is seen only much more weakly at -150 \kms.  CO emission is detected at -22 \kms\
and with low significance at -150 \kms,\ but not at $-190$ \kms.  CO emission at high 
negative velocities is, however, present in the vicinity of B1741-312. 
In Fig. 2  we superpose the \hcop\ absorption profile on a latitude-velocity cut 
at l = -2.125\degr\ in CO emission, 
illustrating the peculiar geometry and non-circular kinematics of gas in the bulge.  
CO emission from the bulge, at $|$v$| > 30$ \kms,  appears preferentially above 
the Galactic equator owing to the tip and tilt  of the inner galaxy gas 
\citep{BurLis78,LisBur78} with the high positive-velocity gas on the far 
side of the gas distribution appearing  at higher latitude  
\footnote{CO emission at -54 \kms\ that is centered on the
Galactic equator arises in the so-called near 3 kpc arm \citep{DamTha08}
and does not share the tilted geometry.}. 
CO emission around -190 \kms\ is present in Fig. 2 only at 
$ b > -0.5$\degr\, at a minimum  projected distance of 75 pc.  

The distribution of CO emission is shown more comprehensively in Fig. 3, which
 locates B1741-312 on the sky as seen in slices 20 \kms wide of the CO 
emission data cube.  CO emission is detected at -190 \kms\ within some 
10-15\arcmin\ of the position of B1741-312, a projected separation of 25 - 35 pc.
The molecular absorption at -150 \kms\ is seen just at the periphery of a nearby 
CO emission feature, as also suggested in Fig. 2.
  
\subsection{Placement  of kinematic features}

Gas kinematics in the Galactic disk are dominated by circular rotation
while motions in the Galactic bar and bulge are characterized by strongly 
non-circular velocities \citep{Fux99}.  Figure 4 shows the variation in 
LSR velocity with distance and galactocentric radius for pure circular motion 
with a flat rotation curve with R$_0$ = 8.5 kpc, $\Theta = 220$ \kms, and 
for a non-circular velocity field inside \Rgal\ = 1.5 kpc using
approximately equal circular and radial motions 170-180 \kms\ to explain the 
geometry and gas motions in the inner galaxy, including the stronger
CO emission at higher latitude in Fig. 2 \citep{BurLis78}.  In pure rotation, 
velocities v $<$ -140 \kms\ arise
exclusively within 300 - 500 pc of the center: with non-circular motion the 
velocity field is bifurcated, and velocities v $<$ -140 \kms\ arise at both larger
and smaller radii in the region interior to \Rgal\ = 1.5 kpc (the outermost radius 
of the underlying model).

The line-of-sight velocity gradient for disk gas in pure rotation is quite 
shallow within 5-6 kpc of the Sun, and the LSR velocity at the edge of 
the bulge gas distribution at \Rgal\ = 1.5 kpc in Fig. 4 is -40 \kms.  A separation 
into bulge gas at v $\la -140$ \kms\ and disk gas at v $\ga$ -40 \kms\ works 
quite well for the molecules seen toward B1741-312 and for most of the H I
because there are no features or strong emission in the intermediate velocity 
range.  H I emission from the 3 kpc arm is seen in the intermediate velocity 
range, and the 3 kpc arm is usually placed in the transition region between 
the bulge and disk gas distributions \citep{Fux99}.

\subsection{Derived results}

To provide a quantitative assessment of the importance of the molecular absorption,
Table 1 summarizes the observed line properties for a set of velocity ranges framing 
the features that were detected in \hcop.  The table entries from the top down are as 
follows:

\begin{itemize}

\item{\Rgal: galactocentric radius. For the two most negative velocity ranges, the 
 gas is ascribed to the bulge at \Rgal\ = 320 - 1500 pc without greater specificity
 given the bifurcation in the velocity shown in Fig. 4 and discussed in Sect. 3.1:
 320 pc is the pericenter distance.
 The other ranges of galactocentric radius were  assigned on the basis of the 
 flat rotation curve with R$_0$ = 8.5 kpc, $\Theta$ = 220 \kms.  The identical
radius ranges given for the two highest velocity intervals both correspond to 
v $\ge -15$ \kms\ in pure rotation.  The molecular gas seen at 7 \kms\  must be 
near the Sun, given the strong 
H I self-absorption, even though positive velocities are not
allowed by rotational kinematics within the solar circle toward B1741-312.
H I emission at v $\ga 20$ \kms\ probably arises in the inner galaxy at an uncertain
galactocentric radius, and its contribution to the column density, 7.3\% of the
total, has been ignored in this table.}


\item{W$_{\rm X}$: the integrated observed optical depth for HCN, \hcop,\ and HNC (units
of \kms) and integrated brightness temperature for H I and CO observed in emission,
in units of K-\kms.}

\item{N(X): column densities (units of $\pcc$) calculated using Eq. 1 for H I,
\Texc\ = 2.729 K for \hcop, and with N(\hcop)/N(\HH) $= 3 \times 10^{-9}$, the local
value for solar metallicity as a benchmark  for \HH.  A  fixed
conversion taken from the values determined near the Sun was also used by 
\cite{SchChe+14}, whose 3D E(J-K) extinction maps we used to derive the variation
of N(H) as inferred from dust measures.}

\item{X$_{\rm CO}$:  N(\HH)/\WCO, the CO-\HH\ conversion factor.  We note that the assumed
\hcop\ abundance yields a canonical value $2\times 10^{20}$ \HH$\pcc$ (K-\kms)$^{-1}$
for the CO-\HH\ conversion factor for the two 
velocity intervals at v $\ge -15$ \kms, independent of any assumption about the 
rotation curve or placement of the gas in the Galaxy. \cite{SchChe+14} derived a
 similar CO-\HH\ conversion factor in the vicinity of B1741-312 when N(H) was derived 
from the total E(J-K) and compared with  profile-integrated total CO measurements from 
the NANTEN2 telescope.  This occurs because the gas column is dominated by the local 
material seen around zero-velocity.} 


\end{itemize}

\begin{figure*}
\includegraphics[height=14cm]{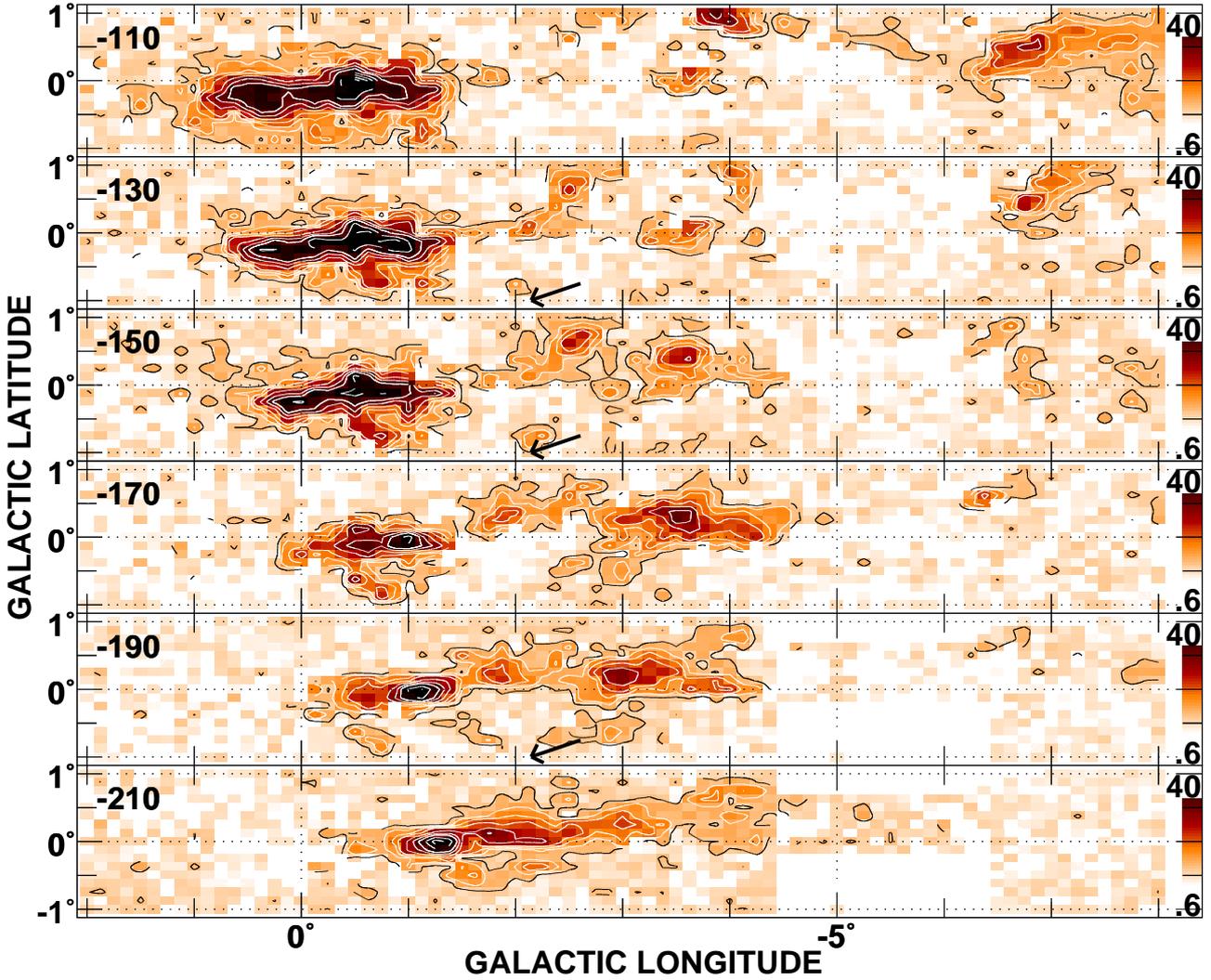}
\caption{CO integrated intensity maps in 20 \kms\ intervals centered as indicated
in each panel.  An arrow marks the position of B1741-312
at $l=-2.1366$\degr,  $b=-0.9968$\degr\ in panels where molecular
absorption was detected.  The color scale 
extends from 0.5 to 40 K- \kms.  Contours are shown at 2, 4, 7, 11, 20, 
30, 40, 60, 80, 100, and 120 K- \kms.}
\end{figure*}

The tabulated values of N(H I) and N(\HH) show a marked change from an approximately 
equal mixture of atomic and molecular gas in the disk at \Rgal\ $>$ 3 kpc to a predominantly 
molecular medium in the bulge at \Rgal\ $<$ 1.5 kpc.  Deriving the \HH\ column density from CO 
emission using a standard CO-\HH\ conversion factor 
N(\HH)/\WCO\ = $2\times 10^{20}\pcc$ (K-\kms)$^{-1}$ \citep{BolWol+13} also yields approximately
equal proportions locally,  but CO emission underrepresents the molecular gas at 
\Rgal\ $<$ 1.5 kpc. This is in strong contrast to the low values for the CO-\HH\ conversion 
factor that apply in the CMZ where CO is overluminous \citep{BolWol+13,SchChe+14}.  The transition region
at \Rgal\  = 1.5 - 3 kpc represented by the velocity range -41 \kms $\le$ v $\le$ -15 \kms\
is predominantly atomic when N(\HH) is derived from \hcop, but with relatively bright
CO, leading to a small implied CO-\HH\ conversion factor.  The proportion of
molecular gas  would be much larger and about equal to that of H I if the
standard local  CO-\HH\ conversion were applied.

\subsection{Importance of the molecular absorption}

In Fig. 5 we show the variation of cumulative total column density N(H) working outward
to the solar circle, as  derived under various assumptions.  The solid curve results from 
interpolating between the binned entries in the 3D extinction data cube of \cite{SchChe+14} 
at $l=-2.1$\degr, $b=-1$\degr\ using a constant scaling 
N(H) $= 1.171\times 10^{22}\pcc$ E(J-K), as discussed in Sect. 2.3.  The extinction 
data do not extend beyond a heliocentric distance of 10.5 kpc, but this should include 
the entire bulge contribution, while only a very small contribution is expected from the 
far-side disk gas at the latitude of B1741-312.  Even the radio spectra do not sample 
much material outside the solar circle, which is seen at z-heights exceeding 300 pc 
on the far side of the Galaxy.

We also show in Fig. 5 values for N(H I) and N(H) where the latter is derived
either using a constant abundance X(\hcop) $ = 3\times 10^{-9}$ or a canonical
CO-\HH\ conversion factor N(\HH)/\WCO\ = $2\times 10^{20} \HH \pcc$ (K-\kms)$^{-1}$.  
The value plotted is integrated over the full corresponding velocity range.  We 
do not pretend to know where within the range the matter is distributed.

The least uncertain values are those for N(H I), so that if the radii are properly assigned,
it is apparent that the extinction requires an additional contribution whose 
magnitude is like that which we have seen in molecular absorption.  In the outer regions 
of the disk, the existence of the molecular gas can be inferred from CO emission using 
a standard CO-\HH\ conversion factor, but this is not the case in the bulge at
\Rgal\ $<$ 1.5 kpc.

\subsection{High \HH-fractions in the bulge gas}

Features associated with the bulge have a remarkably high fraction of hydrogen in 
molecular form, $\sim 0.8$ and $\sim 0.95$ for the -190 and -150 \kms\ velocity 
components, respectively.  Using the analytical framework for the H I/H$_2$ transition 
developed by \cite{sternberg:14} and \cite{bialy:16}, we
derived the parameter $\frac{I_{\rm UV}}{n_{100}}$ where $I_{\rm UV}$ is
the intensity of the FUV radiation field in units of the Draine field and
$n_{100}=\frac{n(H)}{100 \rm cm^{-3}}$ is the gas density in units of
100 H-nuclei $\pccc$.  We find $\frac{I_{\rm UV}}{n_{100}} \sim 0.5$ and $\sim 0.2$ for these two
clouds, indicating that they are exposed to a low far-ultraviolet
(FUV) radiation field. We
note that the density must be modest,  n$_{100} \leq 1$, since no significant 
CO emission is detected. 

The theory indicates that \HH\ is mostly protected by self-shielding in such an
environment, with extinction contributed by dust grains providing
a minor contribution to the protection of H$_2$ against photodissociation.
It is known from  analysis of the spatial distribution of star-forming regions 
in the inner galaxy \citep[e.g.,][]{churchwell:09} that star formation is weak 
in the Galactic bulge, possibly as a consequence of the non-circular motions 
induced by the bar.  The old stellar population of the bulge is expected to produce 
a radiation field dominated by visible and near-IR radiation, which can heat the 
dust \citep{robitaille:12} but is less effective at dissociating molecular hydrogen.
The diffuse gas in the bulge therefore follows the general trend of
matter in the disk, which has an increasing fraction of hydrogen in molecular
form at smaller \Rgal, more than 80\% at \Rgal $\approx$ 2\,kpc \citep{koda:16}.

\subsection{Influence of metallicity variation}

Of course the tabulated values are affected by uncertainty in the  metallicity in the 
bulge gas and disk gas, and an increased metallicity, if reflected in a higher \hcop\ 
relative abundance and dust/gas ratio,  would lead to smaller N(\HH) inferred
from \hcop\ and N(H) inferred from extinction.  A factor three increase in N(\hcop)/N(\HH) 
would yield nearly equal atomic and molecular gas columns at v = -190 \kms,\ but no 
reasonable increase in metallicity would equalize the \HH/H I ratio  at v = -150 \kms,
which is 37.  We note that \cite{SchChe+14}, from whose work we drew the variation
of E(J-K), also  used a constant gas-reddening ratio to derive N(H) and \WCO.

\subsection{Another estimate of the total atomic gas deficit}

The more recent value of the IRAS dust-inferred optical reddening \EBV\ in the direction 
of B1741 at 6\arcmin\ resolution is \EBV\ = 6.3 mag \citep{SchFin11}, which is somewhat 
beyond the stated range of validity of this estimator.  Converting this into total column 
density with a recent value of the N(H)/\EBV\ ratio \citep{Lis14EBVI,Lis14EBVII,HenDra17} 
yields a high total column density 
N(H) $= 6.3 \times 8 \times 10^{21} = 5.0 \times 10^{22}$ H-nuclei $\pcc$.
The implied gas deficit N(H) - N(H I) $\approx 3.5 \times 10^{22}$ H-nuclei $\pcc$ is
much higher than can be explained by either the \hcop\ absorption or the CO emission. 

\begin{figure}
\includegraphics[height=7.5cm]{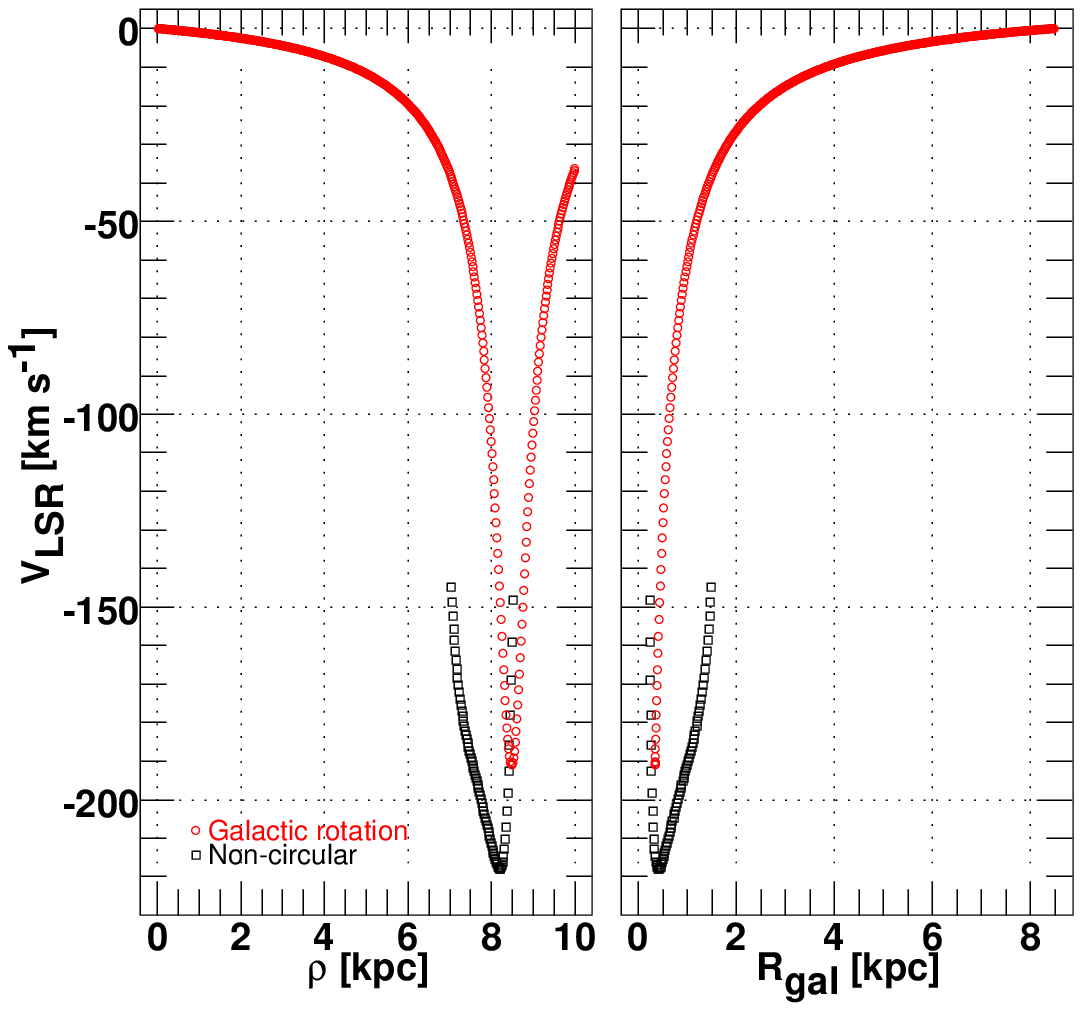}
  \caption[]{Variation of LSR velocity with heliocentric distance (left)
 and galactocentric radius (right) for circular and non-circular velocity
 fields in the direction of B1741-312 at $l=-2.14$\degr, $b=-1.0$\degr.  
 The curves labeled ``Galactic rotation'' are for a flat rotation curve
 with R$_0$ = 8.5 kpc, $\Theta = 220 $ \kms, and the non-circular velocity field
 interior to \Rgal\ = 1.5 kpc is taken from \cite{BurLis78}.  The diagrams were constructed
 by taking equal steps in heliocentric distance, and the density of points represents
 the line-of-sight geometry and velocity gradients.}
 \end{figure}

\begin{figure}
\includegraphics[height=7.5cm]{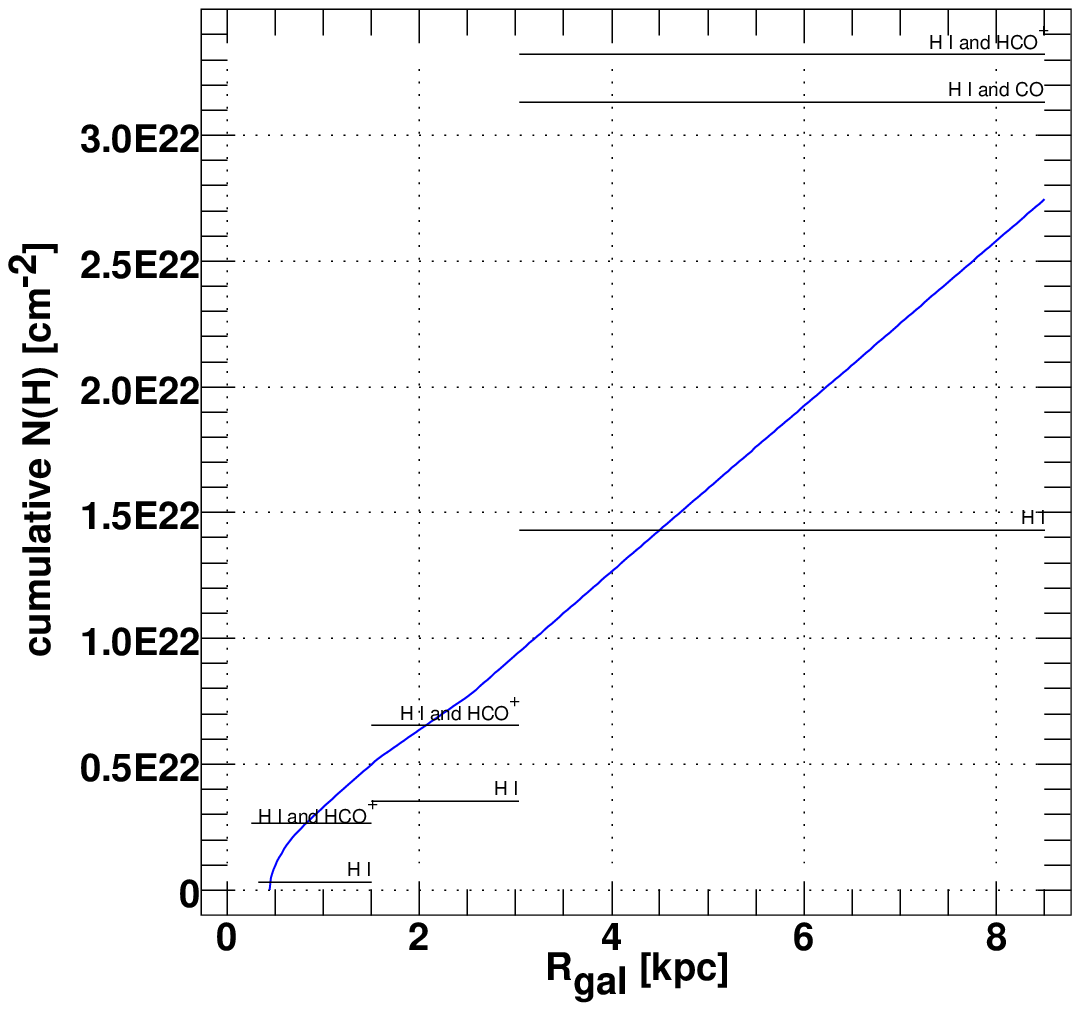}
  \caption[]{Variation in cumulative hydrogen column densities 
N(H I) and N(H) = N(H I) + 2N(\HH) vs. galactocentric radius.  The solid blue curve 
for the total column density N(H) was 
 derived by inverting the variation of E(J-K) with distance modulus in the 3D 
extinction data cube of \cite{SchChe+14}, as  discussed in Section 
 2.3.  The binned estimates for N(H) from H I and \hcop\ or H I and CO assume 
N(\hcop)/N(\HH) $= 3 \times 10^{-9}$ or N(\HH) $= 2 \times 10^{20}\pcc$\WCO. 
The bins  correspond to the ranges of galactocentric radius listed in 
Table 1.}
 \end{figure}

\begin{table*}
\caption[]{Observed and derived quantities$^a$}
{
\small
\begin{tabular}{lccccc}
\hline
     & & & Velocity Range ( \kms)  & \\
Quantity &-196,-180  & -164,-135  & -41,-15 & -15,-1  & -1, 19  \\ 
\hline
\Rgal\ [pc] & 320 - 1500 & 320 - 1500 & 1500-3035 &3035-8500 & 3035-8500 \\
\hline \hline
W$_{\rm H I}$ [K \kms]   & 79.9(3.1) & 32.8(3.8) & 1062(6.1) & 1524(7.6) & 2622(10.4) \\ 
W$_{\hcop}$ [ \kms]& 0.802(0.025) & 3.381(0.036) & 1.108 (0.031) & 9.10(0.03)  & 12.22(0.32) \\
W$_{\rm HCN}$  [ \kms] & 0.640(0.027) & 1.310(0.036) & 0.444(0.032)  &8.62(0.03) & 9.65(0.03) \\ 
W$_{\rm HNC}$  [ \kms] &              & 0.333(0.034) & 0.062(0.031) & 1.42(0.03) & 1.72 (0.03) \\
\WCO\  [K \kms]   & 0 (0.44)      &1.40(0.69)   & 5.55(0.50) & 14.53(0.42) & 22.37(0.49) \\
\hline \hline
N(H I) [$\pcc$] & $1.51(0.06)\times10^{20}$ & $6.13(0.71)\times10^{19}$  & $2.77(0.02)\times10^{21}$  
 & $4.05(0.02)\times10^{21}$  & $6.16(0.02)\times10^{21}$ \\
N(\hcop) [$\pcc$] &  $8.58(0.26)\times10^{11}$  & $ 3.62(0.03)\times10^{12}$  &  $1.18(0.03)\times10^{12}$&  
 $9.74(0.28)\times10^{12}$  &  $1.31(0.03)\times10^{13}$ \\
N(\HH) [$\pcc$] &  $2.86\times10^{20}$  &  $1.21\times10^{21}$  & $3.92 \times10^{20}$&  
$ 3.25\times10^{21} $& $ 4.36 \times10^{21}$ \\
\hline \hline
X$_{\rm CO}$[$\pcc$/K- \kms]   & $>$ 2.17 $\times10^{20}$$^b$ & $8.64\times10^{20}$ & 
$7.06\times10^{19}$ & $2.24\times10^{20}$ &$ 1.95\times10^{20}$ \\
\hline
\end{tabular}}
\\
$^a$ See Sect. 2.3 for a description of table entries \\
$^b$ Using the $3\sigma$ upper limit for \WCO\ \\
\end{table*}

\section{Summary}

We began by noting the accidental observation of mm-wave \hcop, HCN and HNC  absorption at 
-150 \kms\ and -190 \kms\ toward the extragalactic mm-wave continuum background source 
B1741-312 seen at l = -2.1366\degr, b = -0.9968\degr\ (Fig. 1).   There is weak CO emission 
in the vicinity of the position of B1741 (Figs. 2 and 3),  but prominent CO emission at such 
velocities is  confined to smaller longitudes and latitudes nearer the Galactic equator.

Emission and absorption at velocities v $\la -150$ \kms\ cannot be precisely located along 
the line of sight owing to imprecise knowledge of the bar-like inner-galaxy velocity field, 
but must arise in gas located within the Galactic bulge (Fig. 4).  Similarly, strong disk 
absorption and 
emission at -40 \kms\ $\le$ v $\le$ 0 \kms\ can be understood in terms of a flat disk rotation 
velocity curve with R$_0$ = 8.5 kpc, $\Theta$ = 220 \kms, but the bright emitting gas at 
0 \kms\ $\le$ v $\le$ 20 \kms\ is not permitted by rotational kinematics and cannot 
be precisely located either, although we know that it must be nearby because it is self-absorbed 
in H I.

The atomic and molecular absorption profiles were interpreted on this basis in Table 1, 
where we assigned  broad ranges in galactocentric radius to the kinematic intervals 
containing the observed molecular features. We derived column densities  N(H I), 
N(\hcop), and N(\HH) assuming as a baseline the locally-determined value 
N(\hcop)/N(\HH) $ = 3\times 10^{-9}$.  This \hcop-based estimate of 
N(\HH) yields canonical values of the CO-\HH\ conversion factor 
N(\HH)/\WCO\ $ = 2.0-2.2 \times 10^{20} \HH\ \pcc$ (K-\kms)$^{-1}$ for
either of the velocity intervals around 0 \kms\ in Table 4. 

Broadly speaking, the balance between the atomic and molecular gas constituents shifts 
from approximate equality in the disk to a mostly molecular medium within the bulge. 
Figure 5 shows  the galactocentric radial distribution of total column density
N(H) derived by inverting the run of IR reddening E(J-K) with distance modulus in 
the 3D extinction data cube of \cite{SchChe+14} using a fixed
relationship  N(H) $= 1.171\times 10^{22}\pcc$ E(J-K) as described in Sect. 2.3.  
H I  can account for only a small portion of the required gas column inside 
the bulge at \Rgal\ $< 1.5$ kpc.  The gas shortfall can be explained by the 
observed \hcop\ absorption using the locally determined fractional abundance 
N(\hcop)/N(\HH) $= 3\times 10^{-9}$ , but CO emission is weak or absent and would 
require a CO-\HH\ conversion factor $\approx 10^{21}$ \HH\ $\pcc$ (K-\kms)$^{-1}$.
This stands in marked contrast to the overluminous CO emission from the central
molecular zone where the N(\HH)/\WCO\ ratio is believed to be 3-10 times lower 
than the  value near the Sun.

The physical properties and total mass of the molecular gas in the bulge remain to 
be determined by observation.  We suggested that the atomic-molecular transition 
between the disk and  bulge results from a weak UV radiation field inside the bulge 
with its older stellar population, encouraging \HH\ formation at relatively low 
number and column densities in otherwise atomic gas.  Suppression of CO emission 
at brightness levels of 1-2 K (typical of
existing survey detection limits) would require densities that are characteristic
of mostly atomic H I clouds in the disk, n(H) $\la 30-50 \pccc$.  Information on the 
in situ number density in the absorbing molecular gas can be derived by 
observing \hcop, HCN, or HNC in emission around the position of B1741-312, given 
the column densities measured in absorption and the fact that the emission
brightness at the limit of detectability varies only as the n(H)-N(X) product
 \citep{LisPet16}.

 ALMA Cycle 4 observations of a small handful of other suitably strong mm-wave 
continuum sources are underway, but none are as fortuitously positioned at such 
small longtitude and latitude as B1741-312.
We note that even if the sort of detailed discussion presented here cannot 
currently be extended to other sightlines, existing 3D measurements of the 
reddening can be compared to H I emission at bulge velocities to determine whether 
the atomic gas deficit seen here is a general phenomenon, as we would predict.

\begin{acknowledgements}

This paper makes use of the following ALMA data: ADS/JAO.ALMA\#2013.1.01194.S .
ALMA is a partnership of ESO (representing its member states), NSF (USA)
and NINS (Japan), together with NRC (Canada), NSC and ASIAA (Taiwan), and
KASI (Republic of Korea), in cooperation with the Republic of Chile.  The
Joint ALMA Observatory is operated by ESO, AUI/NRAO and NAOJ.
The National Radio Astronomy Observatory is a facility of the National Science 
Foundation operated under cooperative agreement by Associated Universities, Inc. 

This work was supported by the French program “Physique et Chimie du Milieu 
Interstellaire” (PCMI) funded by the Conseil National de la Recherche Scientifique 
(CNRS) and Centre National d’Etudes Spatiales (CNES).      

We thank Edwige Chapillon and Philippe Salom\'e for help with the ALMA data processing 
and the referee for helpful comments.  HSL is grateful to the ENS and FREI for
their hospitality during the completion of this manuscript.

\end{acknowledgements}

\bibliographystyle{apj}

\end{document}